\documentclass[11pt,letterpaper]{article}

\usepackage{times}
\usepackage{relsize}
\usepackage{microtype}
\usepackage{xspace}
\usepackage[table,xcdraw,dvipsnames]{xcolor}
\usepackage{color,colortbl}
\usepackage{subfigure}
\usepackage{alltt}
\usepackage{url}
\usepackage{comment}
\usepackage{graphicx}
\usepackage{relsize}
\usepackage{paralist}
\usepackage{todonotes}
\usepackage{wrapfig}
\usepackage{titlesec}
\usepackage{epsfig}
\usepackage{geometry}
\usepackage{tikz}
\usepackage[normalem]{ulem}
\usepackage{booktabs}
\usepackage{wrapfig}
\usepackage{multirow}
\usepackage{pifont}
\usepackage{wrapfig}
\usepackage{soul}
\usepackage{enumitem}
\usepackage{tcolorbox}
\usepackage{pdfpages}
\usepackage{fancyhdr}
\usepackage{longtable}
\usepackage{makecell}
\usepackage{lettrine}
\usepackage{listings}
\usepackage{longtable}
\usepackage[numbers]{natbib}
\usepackage{wrapfig}

\AtBeginDocument{
  \definecolor{pdfurlcolor}{HTML}{097969}
  \definecolor{pdfcitecolor}{rgb}{0,0.6,0}
  \definecolor{pdflinkcolor}{HTML}{097969}
  \definecolor{light}{gray}{.85}
  \definecolor{vlight}{gray}{.95}
}
\usepackage[colorlinks=true,citecolor=pdfcitecolor,urlcolor=pdfurlcolor,linkcolor=pdflinkcolor,pdfborder={0 0 0}]{hyperref}

\titlespacing{\section}{0pc}{1pc}{0.5pc}
\titlespacing{\subsection}{0pc}{1pc}{0.5pc}
\titlespacing{\subsubsection}{0pc}{0.5pc}{0.5pc}
\titleformat*{\section}{\Large\bfseries}
\titleformat*{\subsection}{\large\bfseries}
\titleformat*{\subsubsection}{\normalsize\bfseries}

\lstdefinestyle{wcsStyle}{
  columns=fullflexible,
  tabsize=2,
  showspaces=false,
  showstringspaces=false,
  aboveskip=0em,
  belowskip=0em,
}

\definecolor{mypurple}{HTML}{edf8da}
\definecolor{ruler}{HTML}{228B22}

\definecolor{Gray}{gray}{0.9}

\newcommand{\pp}[1]{\medskip \noindent \textbf{\emph{#1.}}\xspace}

\let\oldheadrule\headrule% Copy \headrule into \oldheadrule
\renewcommand{\headrule}{\color{ruler}\oldheadrule}% Add colour to \headrule

\DeclareUrlCommand\url{\color{ruler}}

\usetikzlibrary{arrows.meta,positioning,calc,backgrounds,fit,shapes.geometric}

\begin{document}

% Headers
\pagestyle{fancy}
\fancyhf{}
\lhead{\color{gray}\smaller \uppercase{Toward Trustworthy Autonomous Science: A Two-Year Community Roadmap}}
\rfoot{\thepage}

% Cover
\includepdf[pages=-]{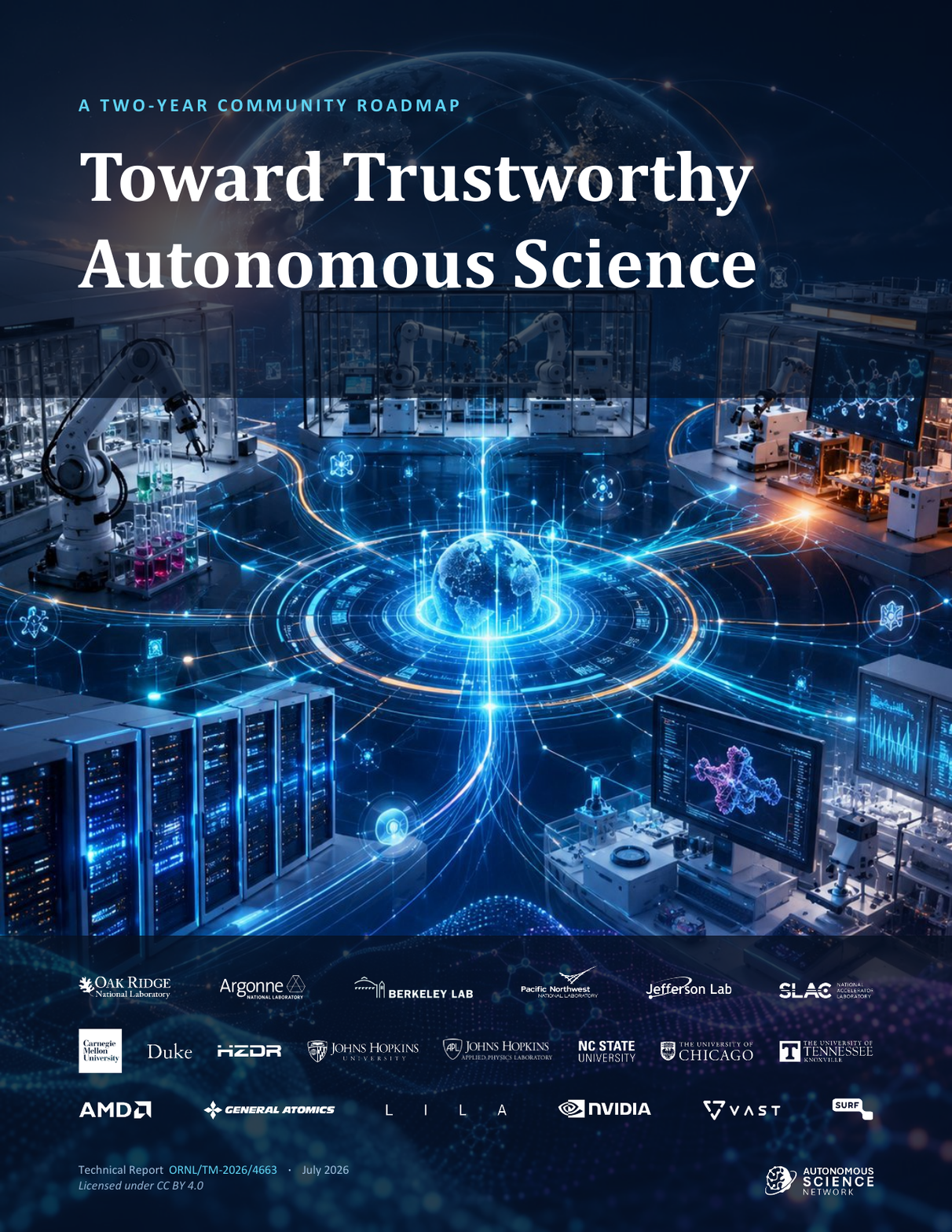}

% Front matter
%%%%% Disclaimer / License %%%%%
\begin{table}[!ht]
\centering
\smaller
\begin{tabular}{p{16cm}}
    \textbf{Disclaimer.}
    This report describes products of research sponsored by the U.S. Department of Energy, Office of Science, Office of Advanced Scientific Computing Research under Contract No. DE-SCL0000175, ``A Testbed for Multi-Agent Autonomous Science: From Lab Bench to Supercomputer".
    \\
    \vspace{-0.25em}
    This report was prepared as an account of work sponsored by agencies of the United States Government. Neither the United States Government nor any agency thereof, nor any of their employees, makes any warranty, express or implied, or assumes any legal liability or responsibility for the accuracy, completeness, or usefulness of any information, apparatus, product, or process disclosed, or represents that its use would not infringe privately owned rights. Reference herein to any specific commercial product, process, or service by trade name, trademark, manufacturer, or otherwise, does not necessarily constitute or imply its endorsement, recommendation, or favoring by the United States Government or any agency thereof. The views and opinions of authors expressed herein do not necessarily state or reflect those of the United States Government or any agency thereof.
    \\
    \vspace{-0.25em}
    \textbf{License.}
    This report is made available under a Creative Commons Attribution 4.0 International Public license ({\small \url{https://creativecommons.org/licenses/by/4.0}}).
\end{tabular}
\end{table}

\vspace{-0.5em}
%%%%% Preferred citation %%%%%
\begin{table}[!ht]
\centering
\smaller
\begin{tabular}{p{16cm}}
    \textbf{\small Preferred citation}
    \\
    R.~Ferreira da Silva, M.~Abolhasani, P.~Beaucage, L.~Biven, M.~Bussmann, K.~Chard, R.~Coffee, S.~DeWitt, S.~Dolas, C.~Eckert, D.~Elbert, I.~T.~Foster, T.~Ghosal, A.~Giannakou, T.~Gibbs, L.~Hamilton, G.~Lockwood, T.~Mayer, B.~Mintz, R.~Nazikian, S.~Nimer, A.~Randles, W.~Shin, S.R.~Sukumar, F.~Suter, M.~Taheri, M.~Taufer, D.~Vrabie, ``\emph{Toward Trustworthy Autonomous Science: A Two-Year Community Roadmap}'', Technical Report, ORNL/TM-2026/4663, July 2026.
    \\
    \rowcolor[HTML]{F7F7F7}
    \lstset{basicstyle=\scriptsize,style=wcsStyle}
    \begin{lstlisting}
@techreport{autonomousscience2026roadmap,
  author      = {Ferreira da Silva, Rafael and Abolhasani, Milad and and Beaucage, Peter and Biven, Laura and Bussmann, Michael and Chard, 
           Kyle and Coffee, Ryan and DeWitt, Stephen and Dolas, Sagar and Eckert, C. and Elbert, David and Foster, Ian T. and Ghosal, 
           Tirthankar and Giannakou, A. and Gibbs, Tom and Hamilton, Leslie and Lockwood, Glenn and Mayer, Theresa and Mintz, 
           Benjamin and Nazikian, Raffi and Nimer, Salahudin and Randles, Amanda and Shin, Woong and Sukumar, Sreenivas Rangan 
           and Suter, Fr\'ed\'eric and Taheri, Mitra and Taufer, Michela and Vrabie, Draguna},
  title       = {{Toward Trustworthy Autonomous Science: A Two-Year Community Roadmap}},
  year        = {2026},
  number      = {ORNL/TM-2026/4663},
  institution = {Oak Ridge National Laboratory}
}
    \end{lstlisting}
    \\
\end{tabular}
\end{table}

\newpage

% Rxecutive Summary
\section*{Executive Summary}

\lettrine[lines=3, findent=2pt, nindent=4pt]{\textcolor{ruler}{\textbf{A}}}{ year ago}, we argued that autonomous laboratories remained \emph{isolated islands}, and we proposed a grassroots network, the \textbf{Autonomous Interconnected Science Lab Ecosystem (AISLE)}, organized around five critical dimensions for connecting them.
The field has since moved faster than that roadmap anticipated.
Multi-agent systems have produced \textbf{experimentally validated hypotheses}; self-driving laboratories have grown markedly more \textbf{interoperable}, increasingly coordinated by shared orchestration software; \textbf{AI agents}, built on reasoning-trained models and scientific foundation models, have grown markedly more capable of multi-step reasoning and tool use; and a national mobilization, the \textbf{Genesis Mission}, has placed autonomous experimentation at the center of U.S. federal science strategy.
Most strikingly, \textbf{industry has become a primary actor}, as well-capitalized entrants build robotic discovery laboratories at a pace that public programs cannot match.

Progress has encountered a sobering counter-current.
A published \textbf{correction} to a flagship autonomous-discovery result retracted its novelty claims and removed a training-data leak; a growing body of \textbf{benchmarks} shows that agents that rival experts on closed-ended questions still complete only a small fraction of open-ended research; and fabricated citations have surfaced even in papers accepted at leading venues.
Tellingly, no system has yet \emph{made and experimentally self-validated a genuinely novel discovery from end to end}.
We read this not as a transient growing pain but as the \emph{defining tension} of the field.

\begin{tcolorbox}[colback=mypurple, colframe=ruler, boxrule=1pt, arc=3pt, left=10pt, right=10pt, top=5pt, bottom=5pt]
\centering
{\large\bfseries\color{ruler!70!black} Producing a candidate discovery is no longer the hard part. Verifying it is.}\\[3pt]
{\normalsize This asymmetry, more than raw model capability, is what limits autonomous science today.}
\end{tcolorbox}

Accordingly, this report organizes the roadmap around \textbf{seven dimensions}, the five we revisit and the two we elevate from cross-cutting afterthoughts to first-class dimensions.

\begin{itemize}[leftmargin=1.6em, itemsep=1.5pt, topsep=3pt, parsep=0pt, label=\textcolor{light}{\textbullet}]
  \item \textbf{Instrument and cyberinfrastructure integration.} Vendor-neutral access to increasingly interoperable self-driving laboratories, with AI inference moving onto the instruments themselves.
  \item \textbf{Agent-driven data management.} Provenance and quality enforced at the point of capture, where scarce data are made AI-ready by construction.
  \item \textbf{Agent-driven orchestration.} Reasoning agents that plan and coordinate specialized scientific methods, and that explain why a result holds.
  \item \textbf{Interoperable agent interfaces.} The consolidating MCP and A2A stack, now joined by an emerging skills layer.
  \item \textbf{Education and workforce development.} The judgment to resist the homogenizing pull of automation.
  \item[\textcolor{ruler}{\ding{72}}] \textbf{Trust, verification, and reproducibility.} \textcolor{ruler!75!black}{\emph{(elevated)}} Verification and validation across the lifecycle as a first-class requirement.
  \item[\textcolor{ruler}{\ding{72}}] \textbf{Safety, security, integrity, and governance.} \textcolor{ruler!75!black}{\emph{(elevated)}} Screening and accountability where digital design meets physical execution.
\end{itemize}

For each dimension, we assess the original milestones (\textbf{M1--M14}), classifying each one as achieved, partially achieved, reframed, or open, and we add \textbf{four new milestones} (\textbf{M15--M18}) for the two elevated dimensions.
We scope the path forward to a \textbf{two-year horizon}, with the first year concentrating on interfaces, protocol adoption, and the scaffolding of verification, and the second targeting federation, zero-trust coordination, and governance.
Throughout, we position the grassroots network as the \emph{interoperability fabric} that allows national programs, international initiatives, and commercial platforms to connect rather than re-silo.

\newpage

% Table of contents
\tableofcontents
\newpage

% Body
\section{Introduction}
\label{sec:intro}

The scientific discovery process is being reshaped by the convergence of automation, robotics, machine learning (ML), and artificial intelligence (AI).
Modern instruments generate data at rates that outpace human analysis, and the cadence of human decision-making is increasingly mismatched with the speed at which experiments can be planned, executed, and interpreted~\cite{silva2025aisle}.
Autonomous science addresses this mismatch by closing the loop between hypothesis, experiment, and analysis, with the goal of compressing discovery cycles that once took years or decades into months or weeks.

One year ago, we argued that this promise was being held back by fragmentation, in that autonomous laboratories operated as isolated islands, unable to communicate across institutional or disciplinary boundaries~\cite{silva2025aisle}.
To address this, we proposed the Autonomous Interconnected Science Lab Ecosystem (AISLE), a grassroots network organized around five critical dimensions: (1)~instrument and cyberinfrastructure integration, (2)~agent-driven data management with FAIR compliance~\cite{fair}, (3)~agent-driven autonomous orchestration, (4)~interoperable agent interfaces, and (5)~education and workforce development.
For each dimension, we surveyed the state of the art, identified open challenges, and proposed a set of milestones (M1--M14) to guide community efforts.

The intervening year has been unusually eventful, and on balance, it has moved faster than the original roadmap anticipated.
Multi-agent systems built on large language models (LLMs) have generated hypotheses that were subsequently validated in the laboratory, from {SARS-CoV-2} nanobody design~\cite{virtuallab} to drug-repurposing candidates in oncology and ophthalmology~\cite{coscientist,robin}.
The interface ``plumbing'' that the original roadmap called for has begun to consolidate, with the Model Context Protocol (MCP)~\cite{mcp} and the Agent2Agent (A2A) protocol~\cite{a2a} emerging as complementary standards, and with science-specific middleware layering them onto high-performance computing (HPC) and experimental facilities~\cite{academy,sciencemcp}.
Most consequentially for a roadmap aimed at national-scale coordination, the launch of the Genesis Mission~\cite{genesis} has placed robotic laboratories and autonomous experimentation at the center of U.S. federal science strategy, alongside parallel efforts in other regions~\cite{euraise,accelerationconsortium}.

Progress has been accompanied by a sobering counter-current.
A published correction to a widely cited autonomous materials-discovery result walked back its central novelty claims, re-characterizing ``novel'' compounds as novel to a prediction platform rather than to science, and removing one compound that had leaked from the training data~\cite{alab_correction}.
A growing body of benchmarks shows that agents which approach expert performance on closed-ended scientific questions still complete only a small fraction of open-ended, end-to-end research tasks~\cite{astabench,scienceagentbench,scicode}, and fabricated citations have begun to appear even in papers accepted at leading venues~\cite{fabricatedcites}.
We claim that these are not transient growing pains but the defining tension of the field.
\textbf{We can now generate candidate discoveries faster than we can verify them.}
Consequently, an updated roadmap cannot treat trust, verification, safety, and governance as cross-cutting concerns folded into other dimensions; they must become first-class dimensions of the roadmap itself.

In this report, we present an updated community roadmap for interconnected autonomous science, one year after AISLE, scoped deliberately to a two-year horizon.
Given how quickly the field is moving, we favor a two-year roadmap over a longer-range vision, so that the milestones we propose remain concrete and accountable rather than speculative. We group them into targets for the first year and targets for the second.
This work makes the following contributions:
\begin{enumerate}
  \item We characterize how the landscape of autonomous science has changed over the past year, organized around seven shifts that bear directly on the original roadmap (Section~\ref{sec:state}).
  \item We assess progress against the original AISLE milestones (M1--M14), classifying each as achieved, partially achieved, reframed, or open, and we refine the five original dimensions accordingly (Section~\ref{sec:dims}).
  \item We elevate two concerns to first-class dimensions of the roadmap, namely trust and verification (Section~\ref{sec:trust}), and safety, security, and governance (Section~\ref{sec:governance}), and we propose milestones for each.
  \item We position the grassroots AISLE network within the new federal and international landscape (Section~\ref{sec:ecosystem}), arguing that top-down mobilization and bottom-up coordination are complementary rather than redundant.
\end{enumerate}

Figure~\ref{fig:arch} shows the resulting structure in two complementary views.
Figure~\ref{fig:arch}(a) renders the architecture as a closed discovery loop in which federated AI agents, drawing on a shared data fabric, hypothesize and design, execute on self-driving laboratories and instruments, capture and curate data, and analyze and verify results, with MCP and A2A as the layers that connect agents to tools and to one another, all enclosed by the two dimensions we elevate to first-class, trust and verification on the inside, and safety, security, and governance on the outside, above an education and workforce foundation.
Figure~\ref{fig:arch}(b) recasts the same roadmap as a two-year trajectory that climbs the autonomy ladder from tool to analyst to scientist, grouping representative milestones into successive phases so that trustworthy autonomy rises to meet, rather than outrun, frontier model capability.

\definecolor{cGov}{HTML}{B2182B}   % governance  (red)
\definecolor{cTru}{HTML}{2C6FA6}   % trust       (blue)
\definecolor{cOrch}{HTML}{6A51A3}  % orchestration (purple)
\definecolor{cInst}{HTML}{E6550D}  % instruments (orange)
\definecolor{cData}{HTML}{2E8B57}  % data        (green)
\definecolor{cEdu}{HTML}{8C6D31}   % education   (brown)
\definecolor{cComm}{HTML}{1597A5}  % communication / protocols (teal)
\definecolor{cCore}{HTML}{2D3A4A}  % core        (slate)

\begin{figure}[tbp]
\centering
\tikzset{
  band/.style={rounded corners=10pt, line width=1pt},
  stage/.style={rounded corners=4pt, draw=#1, line width=1pt, fill=#1!12,
                text=black, text width=2.0cm, align=center, minimum height=1.0cm,
                font=\footnotesize\bfseries, inner sep=2pt},
  core/.style={rounded corners=6pt, draw=cCore, line width=1.2pt, fill=cCore,
               text=white, align=center, minimum height=0.95cm, font=\footnotesize\bfseries},
  fdn/.style={rounded corners=5pt, draw=cEdu, line width=1pt, fill=cEdu!15,
              align=center, minimum height=0.7cm, font=\footnotesize\bfseries},
  loop/.style={->, line width=1.3pt, draw=gray!60},
  mcp/.style={->, >={Stealth[length=1.7mm]}, line width=0.9pt, draw=cComm, dashed,
              shorten >=0.5pt, shorten <=0.5pt},
  a2a/.style={<->, >={Stealth[length=1.7mm]}, line width=0.9pt, draw=cComm,
              shorten >=0.5pt, shorten <=0.5pt},
  stepbox/.style={rounded corners=3pt, line width=1pt},
  bull/.style={align=left, font=\scriptsize, inner sep=1pt},
}

%================= ARCHITECTURE =================
\resizebox{\textwidth}{!}{%
\begin{tikzpicture}[font=\footnotesize, >={Stealth[length=2.3mm]}]
% governance (outer) and trust (inner) enclosing bands
\node[band, draw=cGov!75, fill=cGov!6, fit={(-0.1,0.0) (15.6,5.95)}] (gov) {};
\node[anchor=north, font=\small\bfseries, text=cGov] at (7.75,5.85) {Safety, Security \& Governance};
\node[band, draw=cTru!70, fill=cTru!5, fit={(0.4,0.35) (15.1,5.15)}] (tru) {};
\node[anchor=north, font=\small\bfseries, text=cTru] at (7.75,5.25) {Trust, Verification \& Reproducibility};

% education foundation (full-width)
\node[fdn, text width=12.5cm] (edu) at (7.75,0.82) {Education \& Workforce Development};

% federated agents / data fabric (core), as a wide central bar
\node[core, text width=3.6cm] (core) at (7.75,1.95) {Federated AI Agents $+$ Data Fabric};

% discovery-loop stages in a row (each tied to a dimension)
\node[stage=cOrch] (plan) at (2.6,3.5)  {Hypothesize \& Design};
\node[stage=cInst] (exec) at (6.3,3.5)  {Execute on SDLs \& Instruments};
\node[stage=cData] (capt) at (10.0,3.5) {Capture \& Curate Data};
\node[stage=cTru]  (anal) at (13.7,3.5) {Analyze \& Verify};

% forward pipeline plus a feedback arc that closes the loop
\draw[loop] (plan) -- (exec);
\draw[loop] (exec) -- (capt);
\draw[loop] (capt) -- (anal);
\draw[loop] (anal.north) .. controls (13.7,4.9) and (2.6,4.9) .. (plan.north);
\node[font=\scriptsize\itshape, text=gray!70] at (8.15,4.34) {iterative discovery loop};

% agent-to-agent collaboration (A2A): core <-> reasoning stages
\draw[a2a] (core.west) -- (plan.south);
\draw[a2a] (core.east) -- (anal.south);
\node[font=\scriptsize, text=cComm] at (4.45,2.70) {A2A};

% agent-to-tool access (MCP): core -> instrument/data stages
\draw[mcp] (core.north) -- (exec.south);
\draw[mcp] (core.north) -- (capt.south);
\node[font=\scriptsize, text=cComm] at (7.8,2.80) {MCP};
\end{tikzpicture}}

\par\smallskip
{\small\textbf{(a) Interconnected, closed-loop architecture}}

\par\vspace{8pt}

%================= ROADMAP =================
\resizebox{\textwidth}{!}{%
\begin{tikzpicture}[font=\footnotesize, >={Stealth[length=2.3mm]}]
\node[anchor=north, font=\small\bfseries] at (7.6,6.28) {Closing the capability--reliability gap over two years};

% axes
\draw[->, line width=1pt] (0.5,1.1) -- (14.9,1.1) node[below left=1pt and -2pt, font=\scriptsize]{time};
\draw[->, line width=1pt] (0.5,1.1) -- (0.5,5.95);
\node[anchor=south west, font=\scriptsize] at (0.55,5.58) {autonomy $\uparrow$};

% frontier capability ceiling
\draw[dashed, line width=1pt, draw=cGov!70] (0.6,5.55) -- (14.6,5.55);
\node[anchor=east, font=\scriptsize, text=cGov!80] at (14.6,5.80) {frontier model capability};

% autonomy-ladder ticks
\foreach \y/\lab in {2.0/Tool, 3.3/Analyst, 4.7/Scientist}{
  \draw (0.42,\y) -- (0.58,\y);
  \node[anchor=east, font=\scriptsize] at (0.42,\y) {\lab};
}

% staircase steps (near / mid / long term)
\fill[cInst!16, draw=cInst!75, stepbox] (0.6,1.1)  rectangle (5.2,2.6);
\fill[cOrch!16, draw=cOrch!75, stepbox] (5.2,1.1)  rectangle (10.0,3.7);
\fill[cData!18, draw=cData!75, stepbox] (10.0,1.1) rectangle (14.6,5.0);

% trustworthy-autonomy trajectory climbing toward the ceiling
\draw[line width=1.4pt, draw=cTru, ->]
  (0.6,2.6) -- (5.2,2.6) -- (5.2,3.7) -- (10.0,3.7) -- (10.0,5.0) -- (14.6,5.0) -- (14.6,5.45);
\node[anchor=south west, font=\scriptsize, text=cTru] at (10.15,5.10) {trustworthy autonomy};

% phase labels
\foreach \x/\lab in {2.9/{0--8 mo}, 7.6/{8--16 mo}, 12.3/{16--24 mo}}{
  \node[font=\scriptsize\bfseries] at (\x,0.78) {\lab};
}

% milestone bullets per phase
\node[bull, anchor=north, text width=4.2cm] at (2.9,2.5)
  {$\bullet$ Vendor-agnostic interfaces (M1)\\
   $\bullet$ MCP/A2A adoption (M10)\\
   $\bullet$ AI metadata (M5)};
\node[bull, anchor=north, text width=4.2cm] at (7.6,3.6)
  {$\bullet$ Federated data mesh (M6)\\
   $\bullet$ Verification \& validation (M15)\\
   $\bullet$ Zero-trust comms (M11)};
\node[bull, anchor=north, text width=4.2cm] at (12.3,4.75)
  {$\bullet$ Self-discovering networks (M12)\\
   $\bullet$ Agent identity \& governance (M18)\\
   $\bullet$ National framework (M4)};
\end{tikzpicture}}

\par\smallskip
{\small\textbf{(b) From assistance to trustworthy autonomy}}

\caption{The updated AISLE roadmap. (a)~A closed discovery loop, in which federated AI agents over a distributed data fabric hypothesize and design, execute on self-driving laboratories and instruments, capture and curate data, then analyze and verify, is coordinated by two protocol layers, MCP for agent-to-tool access and A2A for agent-to-agent collaboration across institutions, and is enclosed by the two dimensions this paper elevates to first-class, namely trust and verification and safety, security, and governance, over an education and workforce foundation. (b)~Across a two-year horizon, the roadmap climbs an autonomy ladder from tool to analyst to scientist, with representative milestones per phase, so that trustworthy autonomy rises to close the gap with frontier model capability.}
\label{fig:arch}
\end{figure}

The remainder of this report is structured as follows.
Section~\ref{sec:state} characterizes the current state of autonomous science.
Section~\ref{sec:dims} develops the seven dimensions of the roadmap, the five we revisit, and the two we elevate.
Section~\ref{sec:ecosystem} situates the roadmap within national and global initiatives.
Finally, Section~\ref{sec:conc} concludes the report and outlines a revised milestone roadmap.

\section{A Brief State of Autonomous Science}
\label{sec:state}

Before revisiting the roadmap, we characterize how the landscape has changed since the creation of the AISLE grassroots network.
We organize the discussion around seven shifts and adopt two lenses that the community converged on during the past year.

The first is an \emph{autonomy ladder} that distinguishes degrees of agency.
Recent surveys have largely abandoned the binary vision of automated versus not automated in favor of graded taxonomies of autonomy.
A representative formulation distinguishes three levels, \emph{tool}, \emph{analyst}, and \emph{scientist}, according to whether the system executes well-defined tasks under direct supervision, conducts analysis within human-set boundaries, or formulates hypotheses and proposes new lines of inquiry on its own~\cite{autonomyladder,agenticscience}.
The practical value of such a ladder is that it locates a given system, and a given roadmap milestone, at a specific rung rather than asserting wholesale autonomy.

The second is a persistent \emph{capability-reliability gap} that separates what systems can propose from what can be trusted, that is, the gap between performance on closed-ended scientific questions and performance on open-ended research.
We claim that this gap, rather than raw model capability, is the binding constraint on autonomous science today, and we return to it throughout the report.

\subsection{Shift 1: From Isolated Demonstrations to Validated Discoveries}

A year ago, the strongest claims for autonomous discovery rested on self-driving laboratories optimizing within narrow design spaces.
Since then, multi-agent LLM systems have produced hypotheses that were subsequently validated experimentally.
A virtual laboratory of AI agents designed SARS-CoV-2 nanobodies that were synthesized and shown to bind variant targets~\cite{virtuallab}, an AI co-scientist generated drug-repurposing and target-discovery hypotheses confirmed \textit{in vitro} by collaborating laboratories~\cite{coscientist}, and a multi-agent system proposed a therapeutic candidate for an ophthalmic indication that was validated in follow-up assays~\cite{robin}.
These are genuine advances.
In every case, however, the physical experiments were executed by humans, the problems were human-selected, and at least one celebrated result amounted to re-deriving a mechanism that a human laboratory had already established but not yet published.
At the time of writing, there is no verified instance of an agent autonomously making and experimentally self-validating a genuinely novel discovery end to end. The best-funded recent efforts each miss a different aspect of autonomous discovery, as the systems that reason most autonomously still rely on humans to run the physical experiments~\cite{robin}, the systems that synthesize most autonomously have had their novelty claims contested~\cite{alab_correction}, and the systems that close the full loop without human intervention do so only in computational settings with no wet-lab validation~\cite{aiscientist_v2}.
We revisit both the agents that produce such results and the verification their claims still demand in Section~\ref{sec:dims}.

\subsection{Shift 2: Self-Driving Laboratories Enter a Second Generation}

The self-driving laboratory (SDL) community has begun to describe its own trajectory as a move from a first generation of narrow, hand-tuned, poorly interoperable platforms toward what it calls a second generation, or {SDL}~2.0, that is interoperable, orchestrated, safe, and capable of hypothesis generation~\cite{sdl2,sdl_chemrev}.
The concrete advance so far is interoperability, with the other properties still largely aspirational.
By \emph{orchestrated}, the community means coordinated by a software layer that sequences instruments, robots, and computational steps into managed, restartable campaigns rather than hand-scripted one-off runs.
This shift directly concerns the first three dimensions of the original roadmap, which we revisit in Section~\ref{sec:dims}, where we detail the orchestration frameworks and robotic platforms that make it concrete.

\subsection{Shift 3: The Interface Layer Begins to Consolidate}

When the original roadmap was written, agents reached tools and one another through point-to-point and proprietary interfaces, and we described the requirements in generic terms.
Within a year, two complementary standards have emerged.
The Model Context Protocol (MCP) connects a single agent to many tools and data sources, a vertical concern~\cite{mcp}, while the Agent2Agent (A2A) protocol connects agents to one another across vendor and institutional boundaries, a horizontal concern~\cite{a2a}.
For science specifically, federated-agent middleware and protocol adapters now expose HPC, data-movement, and instrument services to LLM agents through these interfaces~\cite{academy,sciencemcp}.
We treat this consolidation as the most concrete update to the original roadmap, even as the pattern by which agents consume these protocols keeps changing, and we develop it in Section~\ref{sec:communication}.

\subsection{Shift 4: Reasoning and Foundation Models as a New Substrate}

The underlying models have also changed.
Reasoning-oriented LLMs trained with reinforcement learning now approach expert performance on graduate-level scientific question answering~\cite{deepseekr1}, and domain foundation models have produced experimentally corroborated designs in materials and structural biology~\cite{mattergen,alphafold3}.
These models are the substrate on which orchestration (Section~\ref{sec:orchestration}) is increasingly built.
They also sharpen the reliability question because their fluency makes unsupported outputs harder, not easier, to detect.

\subsection{Shift 5: Benchmarks Expose the Capability-Reliability Gap}

A wave of benchmarks has made the second framing lens quantitative.
Agents that perform well on isolated, closed-ended tasks complete only a small fraction of open-ended, end-to-end research tasks~\cite{astabench,scienceagentbench}, and research-grade coding benchmarks curated by scientists remain largely unsolved by frontier models~\cite{scicode}.
Reproducibility-focused agent benchmarks report accuracies that, in some settings, fall below random guessing~\cite{corebench}.
Deployment evidence points the same way, as the first systematic study of agents in production finds that teams keep them deliberately bounded and human-supervised, with most running only a handful of steps before a human intervenes and the majority still gated by human evaluation, and with reliability named as the top obstacle~\cite{agentsprod}.
The lesson for the roadmap is that progress should be measured against open-ended, verifiable tasks rather than against exam-style proxies that are rapidly saturating.

\subsection{Shift 6: Trust and Governance Move to the Foreground}

The past year also supplied concrete evidence that trust and governance can no longer be treated as secondary.
A published correction to a flagship autonomous-discovery result re-characterized its novelty claims and removed a training-data leak~\cite{alab_correction}, fabricated citations appeared in papers at leading venues~\cite{fabricatedcites}, analyses found that journal AI-disclosure policies have done little to curb undisclosed AI-assisted writing~\cite{aipolicy}, and the convergence of capable models with remotely accessible laboratories raised dual-use concerns that the existing policy apparatus does not address~\cite{sdl_policy}.
These developments motivate the two dimensions we add to the roadmap in Sections~\ref{sec:trust} and~\ref{sec:governance}.

\subsection{Shift 7: Industry Becomes a Primary Actor}

Finally, the past year changed who builds autonomous science.
A field that had been driven largely by academic and national-laboratory prototypes acquired well-capitalized industrial entrants, and the change is large enough to bear on every dimension of the roadmap.
Startups founded by senior industry researchers and dedicated to autonomous discovery raised sums that dwarf typical academic budgets, with Lila Sciences assembling roughly half a billion dollars to build robotic ``AI science factories''~\cite{lila} and Periodic Labs raising a \$300~million seed round to pursue closed-loop materials discovery~\cite{periodic}.
The major cloud and chip vendors moved in parallel, shipping agentic research platforms~\cite{msft_discovery} and scientific foundation models~\cite{mattergen}, and positioning large-scale compute and physics-based simulation as the substrate for autonomous experimentation~\cite{nvidia_genesis}.
Commercial cloud laboratories matured into a credible delivery model for remote, programmatic experimentation, a development we take up in Section~\ref{sec:instruments}.
We read this influx as a double-edged development.
It supplies capital, engineering, and infrastructure that the academic community cannot match, yet the distance between the capability these entrants claim and the evidence they have published is itself an instance of the capability-reliability gap, since several of the best-funded autonomous-discovery efforts have released little peer-reviewed validation as of this writing~\cite{lila,periodic}.
A roadmap for interconnected science must therefore treat industry as a primary actor while asking of it the same open interfaces, provenance, and verification that it asks of public laboratories.

\subsection{Relation to Prior Roadmaps and Surveys}

The past year also produced a wave of surveys and roadmaps for autonomous science, including taxonomies of autonomy~\cite{autonomyladder}, broad framings of agentic discovery~\cite{agenticscience}, and domain-oriented catalogs of agentic systems~\cite{agentic_survey_gridach}.
These works are valuable as maps of the literature, and we draw on them for the two lenses above.
Our contribution is different in kind.
Rather than surveying the field anew, we update a specific, milestone-bearing roadmap~\cite{silva2025aisle,silva2024labsfuture} against one year of evidence, score its milestones, and revise its structure where the evidence demands.
We believe that this form of accountable revision, in which a community commits to milestones and then publicly assesses them, is a useful complement to survey-style framings and one that the field currently lacks.
We intend it as a recurring practice rather than a one-off, revisiting and rescoring these milestones as the field advances so that the roadmap stays honest about what has and has not been achieved.

\section{Critical Dimensions of the Roadmap}
\label{sec:dims}

We organize the roadmap around seven dimensions.
Rather than enumerate them in isolation, Figure~\ref{fig:arch} situates them within a single closed-loop architecture, where the discovery-loop stages carry orchestration, instruments, and data, the coordinating protocol layers carry the agent interfaces, the enclosing bands carry trust and governance, and the foundation carries education and workforce development.
The first five revisit and update the dimensions of the original AISLE roadmap~\cite{silva2025aisle} in light of the past year, while the last two, trust and verification (Section~\ref{sec:trust}) and safety, security, and governance (Section~\ref{sec:governance}), are elevated here from cross-cutting concerns to first-class dimensions.
For each dimension, we summarize the state of the art, identify the challenges that remain, state research priorities, and record the status of the associated milestones in the scorecard of Table~\ref{tab:scorecard}.
The subsections that follow justify these entries dimension by dimension, and we interpret the overall pattern in the conclusion (Section~\ref{sec:conc}).

\begin{table}[t]
\centering
\caption{Status of the original AISLE milestones (M1--M14) and proposed new milestones (M15--M18), each with a target year (Y1 or Y2) on the two-year roadmap. Statuses are provisional and subject to confirmation against the cited evidence.}
\label{tab:scorecard}
\small
\setlength{\tabcolsep}{4pt}
\begin{tabular}{@{}c p{0.46\columnwidth} l c@{}}
\toprule
\# & Milestone (abbreviated) & Status & By \\
\midrule
M1  & Common instrument interfaces, HAL            & Partial  & Y1 \\
M2  & End-to-end cross-institution workflows       & Partial  & Y2 \\
M3  & Open compute fabric, fault tolerance, digital twins & Open & Y2 \\
M4  & Scalable national instrument framework       & Open     & Y2 \\
M5  & AI-driven metadata and annotation            & Partial  & Y1 \\
M6  & Federated data mesh, FAIR governance         & Open     & Y2 \\
M7  & Near-real-time processing and AI provenance & Partial  & Y2 \\
M8  & Hierarchical LLM orchestration, verification & Partial  & Y1 \\
M9  & Cross-facility knowledge integration         & Open     & Y2 \\
M10 & Standardized cross-vendor agent interfaces   & Reframed & Y1 \\
M11 & Zero-trust, sub-second agent coordination    & Open     & Y2 \\
M12 & Self-discovering agent networks              & Partial  & Y2 \\
M13 & National education consortium                & Open     & Y1 \\
M14 & Virtual labs, human-AI assessment            & Open     & Y2 \\
\midrule
M15 & Verification and validation across the lifecycle & New  & Y1 \\
M16 & Reproducibility and efficiency benchmark     & New      & Y1 \\
M17 & Screening at digital-physical interface      & New      & Y1 \\
M18 & Cross-institution agent identity, governance & New      & Y2 \\
\bottomrule
\end{tabular}
\end{table}

\subsection{Instrument and Cyberinfrastructure Integration}
\label{sec:instruments}

The first dimension concerns how autonomous agents orchestrate diverse experimental equipment and computational resources across institutional boundaries.
Increasingly, the two are inseparable, as instruments, edge devices, HPC, cloud, and digital twins form a single distributed cyber-physical system in which AI models, simulation, and laboratory automation execute as one coordinated workflow whose reliability determines the trustworthiness of the result.
Over the past year, the defining development has been the community's move toward the second-generation self-driving laboratories (SDLs) described in Section~\ref{sec:state}, more interoperable and orchestrated than the narrow, hand-tuned systems that dominated earlier demonstrations~\cite{sdl2,sdl_chemrev}.

\pp{Brief state of the art}
Orchestration software has matured from bespoke scripts toward reusable frameworks, such as MADSci for modular discovery campaigns~\cite{madsci} and ChemOS~2.0 for chemical SDLs~\cite{chemos2}, while mobile and multi-robot platforms now approach human throughput on real synthesis tasks~\cite{mobilechemist}.
Beneath the agent layer, device-control standards such as SiLA~2 provide vendor-neutral, typed instrument communication~\cite{sila2}, and standards bodies have begun to target the interfaces that a modular autonomous laboratory requires~\cite{nist_sdl}.
At the facility scale, programs that connect instruments, robotic laboratories, and HPC into shared workflows have continued to grow~\cite{intersect}.
A complementary commercial route to instrument access has matured in parallel, namely cloud laboratories that expose physical instruments to remote programmatic control, so that an experiment specified as code is executed by robots and technicians at a central facility~\cite{emeraldcloud}.
This model has reached academia through the first university cloud lab~\cite{cmucloudlab}, and it has already been driven by a language-model agent that learned the facility's scripting language from documentation and carried out cross-coupling reactions end to end~\cite{autonomouschem}.
National programs have begun to move this model from isolated facilities toward a networked resource, most concretely a federal test-bed for a network of programmable cloud laboratories linked by shared networking and data standards (Section~\ref{sec:ecosystem})~\cite{nsfpcl}.
A further shift concerns where computation happens relative to the instrument.
As the data rates of upgraded light sources, particle detectors, and radio arrays began to outpace the store-then-analyze pipeline, reaching exabyte-scale annual volumes at the largest facilities~\cite{ska,heproadmap}, machine-learning inference has moved toward the instrument edge so that detector data are processed as they stream rather than after they are written to storage~\cite{fastml}.
Demonstrations include real-time ptychographic reconstruction from a streaming detector on an edge accelerator~\cite{edgeptycho} and machine-learning triggers that decide within microseconds which events to retain~\cite{atlastrigger}.

\pp{Challenges}
The heterogeneity that motivated the original roadmap persists, as most instruments still ship without a standard control interface, retrofitting drivers is labor intensive, and the majority of deployed SDLs operate at the lower rungs of the autonomy ladder of Section~\ref{sec:state}.
The past year added a sharper concern at the instrument level, namely that conclusions drawn from automated characterization can be wrong in ways that human inspection would catch, as the corrected analysis of an autonomous materials laboratory made clear~\cite{alab_correction}.
Organizational barriers compound the technical ones, since intellectual-property and liability questions arise the moment an instrument in one institution is driven by an agent in another.
A deeper constraint is that the hardware has advanced more slowly than the software around it.
Robotics, sample handling, and reconfigurable experimental setups remain comparatively rigid, and where the apparatus cannot be recomposed under program control, the reach of an agentic workflow is bounded by a largely fixed experiment, which optimization and campaign-driven science can accommodate but open-ended discovery cannot.

\pp{Research priorities}
We reaffirm the need for vendor-agnostic hardware abstraction layers and self-describing instruments that expose their capabilities semantically, and we add the validation of automated characterization as a priority.
In effect, this extends the MCP-style capability discovery of the agent interface layer (Section~\ref{sec:communication}) down to the instrument and its data-acquisition system, so that an agent can discover and drive a detector as readily as any other software tool.
This capability sits above the typed device-control interfaces that standards such as SiLA~2 already provide~\cite{sila2}.
We also add modular, agile, and reconfigurable experimental hardware as a priority in its own right, since vendor-neutral software interfaces deliver little if the apparatus beneath them cannot be recomposed.
Recent work pursues exactly this for the structural refinement that the corrected materials result called into question, building a rubric-bounded agent for Rietveld analysis of X-ray diffraction whose scoring rewards credible fits and flags out-of-scope cases rather than forcing a pattern fit~\cite{agentbuild}.
Physics-aware digital twins should be used to test autonomous workflows before they touch physical instruments, as in self-driving laboratories that validate a workcell in simulation before any physical run~\cite{polybot}, and instrument-level outputs should carry the provenance needed to audit downstream claims (Section~\ref{sec:trust}).
We further prioritize real-time, edge-side inference and data reduction so that analysis and on-the-fly experiment steering keep pace with instruments whose output exceeds what centralized pipelines can absorb, and so that raw streams are turned into analysis-ready data at the point of acquisition (Section~\ref{sec:data})~\cite{fastml}.
Above the individual instrument, the cyberinfrastructure half of this dimension calls for an open compute fabric that spans edge, HPC, cloud, and heterogeneous accelerators, exposing vendor-neutral interfaces for model execution, accelerator scheduling, workflow portability, and provenance capture, much as MCP and A2A do for tools and agents (Section~\ref{sec:communication}), so that workflows move across platforms while hardware and software ecosystems evolve independently beneath them~\cite{doe_iri}.
The status of the original milestones M1 through M4 is summarized in Table~\ref{tab:scorecard}, and the national framework envisioned by M4 is now partly scaffolded by federal mobilization~\cite{genesis}.

\subsection{Agent-Driven Data Management}
\label{sec:data}

The second dimension concerns the shift from centralized repositories to distributed systems in which autonomous agents curate, validate, and federate scientific data, enforcing FAIR principles at the point of capture~\cite{silva2025aisle, fair}.
The past year reframed this dimension around trust.
As autonomous campaigns generate data that humans never inspect, provenance and quality assessment become prerequisites for credible discovery rather than after-the-fact bookkeeping.

\pp{Brief state of the art}
Federated data services coordinate movement and processing across many facilities~\cite{globus}, and pass-by-reference systems allow large datasets to be shared without duplication in distributed agent workflows~\cite{proxystore}.
Provenance models such as PROV-O provide a vocabulary for traceability~\cite{provo}, and standards bodies have begun to target the data and knowledge-management gaps specific to autonomous laboratories~\cite{nist_sdl}.
What remains missing is the autonomous enforcement of these capabilities, i.e., agents that curate and annotate data as it is produced rather than leaving it to downstream human stewardship.

\pp{Challenges}
Beyond the format and schema heterogeneity identified previously, four issues have moved to the foreground.
First, autonomous agents must negotiate schema evolution when they encounter new experiment types, without manual intervention.
Second, data quality is now a verification problem because contaminated or leaked data can propagate silently through AI-driven decision chains, as illustrated by a training-data leak in a flagship autonomous-discovery result~\cite{alab_correction}.
Agents, therefore, require mechanisms to assess reliability from the experimental context, rather than treating all data as equally trustworthy.
Third, automation multiplies the volume of raw data, yet the high-quality, well-characterized data needed to train reliable models remains scarce, and how that data is represented can determine whether a model learns the underlying physics at all~\cite{smalldata,datarepresentation}.
Much of what separates reusable data from raw output is well-curated metadata, the theoretical and experimental context, conditions, and provenance that make a measurement findable, interpretable, and trustworthy~\cite{fair}, and capturing it at the moment of acquisition rather than reconstructing it afterward is what keeps data quality tractable as volume grows.
Fourth, the move to online, in-transit data reduction (Section~\ref{sec:instruments}) sharpens the problem because signals that are filtered or summarized at the instrument edge can never be re-examined.
The FAIR lifecycle and its provenance must therefore be preserved through reduction, recording not only what was kept but also what was removed and why, lest reproducibility be quietly lost as the data is reduced.

\pp{Research priorities}
We prioritize federated data-mesh architectures in which each laboratory maintains a node with standardized interfaces and global discovery indices, support for both explicit and implicit schemas so that agents can infer structure from heterogeneous sources, and the embedding of provenance frameworks~\cite{provo} into instrument middleware so that every autonomous decision is traceable across facilities and timescales.
We also prioritize making data \emph{AI-ready} by construction, with readiness levels that track a dataset from raw capture to a form fit for training~\cite{datareadiness}, and community standards for autonomous-laboratory data that play, for machine consumption, the role that FAIR played for sharing~\cite{fair, nistai}.
Data curated to this standard are not only an audit trail but also the substrate for surrogate models that compress expensive experiments and carry materials knowledge toward manufacturing timescales~\cite{materialsaccel,mgi}.
More broadly, as autonomous agents become the dominant consumers of these data, the systems that serve them are better designed to be \emph{agent-first} from the outset, anticipating the high-volume, exploratory access patterns of agents rather than retrofitting interfaces built for human analysts~\cite{agentfirst}.
The status of milestones M5 through M7 is summarized in Table~\ref{tab:scorecard}.

\subsection{Agent-Driven Autonomous Orchestration}
\label{sec:orchestration}

The third dimension is the cognitive core of interconnected autonomous laboratories, namely the agents that navigate scientific decision spaces while remaining aligned with physical and domain knowledge~\cite{silva2025aisle}.
The substrate for this dimension changed markedly over the past year, as reasoning-oriented LLMs and domain foundation models matured into orchestrators that coordinate specialized methods such as Bayesian optimization, uncertainty quantification, and reinforcement learning.

\pp{Brief state of the art}
Multi-agent systems that generate, debate, and evolve hypotheses have produced experimentally validated results~\cite{coscientist}, and agentic tree-search systems now carry an idea from conception to a written manuscript~\cite{aiscientist_v2}.
The models beneath them have improved on two fronts.
Reasoning-trained LLMs approach expert performance on graduate-level science questions~\cite{deepseekr1}, and domain foundation models produce experimentally corroborated designs in materials and structural biology~\cite{mattergen,alphafold3}.
A complementary line of work casts the LLM as an orchestrator of established tools rather than a replacement for them, for example, by exposing expert-designed chemistry tools to a language model~\cite{chemcrow}.

\pp{Challenges}
The probabilistic nature of LLM-based agents remains in tension with the determinism that reproducible science assumes, and it is unclear how to guarantee reproducible outcomes or grounding in physical law, when an orchestrator is non-deterministic, higher-latency, and difficult to verify.
The capability-reliability gap of Section~\ref{sec:state} makes this concrete, since strong performance on closed-ended tasks has not carried over to open-ended research.
The orchestrator is therefore best understood as one component of the ecosystem, rather than a replacement for the established methods it coordinates.
A recurring response is to keep those methods in the loop, combining data-driven learning with physical and chemical constraints and checking an agent's proposals against known laws before they reach an instrument~\cite{agentic_survey_gridach}.
In this view, the probabilistic reasoning of an LLM is an asset for exploration and a liability for commitment, and the design problem is to route each to where it belongs.

\pp{Research priorities}
We reaffirm three thrusts.
The first is hierarchical architectures in which LLM-based agents orchestrate established scientific methods abstracted as actuators.
The second is a verification and validation infrastructure that uses digital twins and formal or symbolic methods to enforce physics-based constraints as hard boundaries, which we develop in Section~\ref{sec:trust}.
The third is distributed, real-time knowledge integration that keeps agents grounded across facilities and long campaigns.
Grounding of this kind rests on a primitive the original roadmap did not name, namely durable agent memory.
We mean policy-bound, provenance-rich, scoped persistence, spanning the working, episodic, semantic, and procedural forms long studied for cognitive agents, that lets an agent carry context, decisions, and lessons across a long campaign instead of rebuilding them inside a prompt, and that is distinct from both the scientific data of Section~\ref{sec:data} and the model weights beneath it~\cite{coala}.
Early systems realize this idea by managing a small in-context working set against external stores, much as an operating system pages physical memory against a larger virtual address space~\cite{memgpt}.
Cutting across all three, the construction of an agent is worth treating as a workflow stage in its own right, so that a scientist authors a durable contract, a version-controlled rubric, a graded curriculum, and a curated knowledge base, that bounds an automated builder and survives the churn of the underlying models. This has been demonstrated for the orchestration of a crystallographic refinement tool~\cite{agentbuild}.
We also hold that an orchestrator should yield understanding, not only predictions, since an explanation of why a design or result holds is what makes an autonomous finding actionable, and interpretability of this kind is increasingly expected of scientific machine learning~\cite{xai_science,xai_materials}.
A further fragility is dependence on the models themselves, since most deployed agents rely on a small set of proprietary frontier models~\cite{agentsprod}, an uneven foundation for reproducible science and one not equally available across regions.
Open-weight models accompanied by clear provenance of their training data are therefore of growing interest, because reproducibility requires knowing what a model learned from, and because scientific sovereignty requires not depending on a single vendor.
Once agents depend on them throughout a campaign, models are best treated as persistent scientific infrastructure, versioned, evaluated, and retired with the discipline applied to instruments and scientific software rather than swapped silently beneath a running workflow.
The status of milestones M8 and M9 is summarized in Table~\ref{tab:scorecard}.

\subsection{Interoperable Agent Interfaces}
\label{sec:communication}

The fourth dimension concerns the interfaces and standards through which autonomous agents reach the tools, data, and capabilities they depend on, and through which they coordinate with one another across institutional and disciplinary boundaries.
This dimension has changed more than any other since the original roadmap, which described its requirements in generic terms because no widely adopted standard yet existed~\cite{silva2025aisle}.
Within a year, a recognizable two-axis stack has emerged.
The Model Context Protocol (MCP) standardizes how a single agent connects to many tools and data sources, while the Agent2Agent (A2A) protocol standardizes how agents discover, authenticate, and delegate to one another across vendor and institutional boundaries~\cite{mcp,a2a}.
We treat these as complementary, vertical and horizontal respectively, rather than competing~\cite{agentprotocols_survey}.
More recently, a third layer has begun to form above these protocols, namely \emph{agent skills}: composable bundles of instructions, scripts, and resources that an agent discovers and loads on demand. While MCP and A2A govern how agents reach tools and one another, skills capture what an agent needs to know to carry out a task~\cite{agentskills}.

\pp{Brief state of the art}
Both protocols moved under neutral governance within months of each other, which lowers the risk of relying on a single vendor for betting infrastructure~\cite{a2a,agentprotocols_survey}.
For science specifically, federated-agent middleware deploys stateful agents across HPC, data, and experimental resources~\cite{academy}, and thin adapters now expose facility services, such as data movement and remote function execution, to LLM-based agents through MCP~\cite{sciencemcp}.
These layers sit above the instrument-control standards of Section~\ref{sec:instruments}~\cite{sila2}, so that a typed instrument interface can be wrapped for agent access without discarding existing control software.
The way agents consume these interfaces is itself in flux.
Loading every tool definition into the context window and routing each intermediate result back through the model scales poorly, in both token cost and error surface, once an agent faces dozens of tools. A growing practice instead has the agent write code that calls the tools and discovers their definitions on demand, which repositions the protocol as a substrate beneath code execution rather than a direct call interface~\cite{mcp_codeexec}.
A further step, pursued in industry though likely beyond our two-year horizon for science, is a runtime that surfaces tools and agents dynamically, exposing each step of a workflow only to the capabilities relevant to its goal and folding discovery, scoping, and least-privilege access into the interface layer itself.
We read this churn not as the demise of any one protocol but as a reason to standardize the interface while leaving the calling pattern free to evolve, which is the layered stance we take in the priorities below.

\pp{Challenges}
The most-cited blocker is identity and delegated authorization at the facility scale because agents must act on a scientist's behalf across institutions without holding overly broad, long-lived credentials~\cite{sciencemcp}.
A second mismatch is structural, as the request-response tool model fits poorly with the stateful, long-running jobs that characterize scientific campaigns~\cite{academy,sciencemcp}.
Semantic interoperability across domains, the provenance of agent-to-agent decisions, and the tension between protocol fragmentation and convergence all remain open.
The security posture of these protocols also trails their adoption, as national-security guidance warns that placing tool descriptions, control flow, and data in one shared context blurs trust boundaries, overlapping contexts can leak state across tasks, and unverified dynamic tool discovery widens the reach of a compromised component~\cite{mcp_nsa}.

\pp{Research priorities}
We prioritize layered protocol architectures that separate networking, message formatting, semantic interpretation, and coordination, agent identity with attribute-based access control designed for multi-institutional collaboration, provenance-carrying messages that record which agent did what with which tool on which data, and self-describing capability discovery validated on multi-facility testbeds.
On identity in particular, the same runtime that scopes which tools a step may see can also scope its authority, issuing signed delegations on demand and limiting their lifetime to a single invocation rather than to the agent as a whole, a pattern commercial systems already implement and one that facility identity-and-access infrastructure will need to grow to support.
Recording the prompts, decisions, and tool calls exchanged between agents is a solved problem in agent-observability products, so the open question for science is not whether to log these interactions but what a scientifically adequate record must contain, one that ties each decision to the data, tools, and model versions behind it and serves reproducibility and cross-facility reuse rather than operational monitoring alone. We develop that provenance model, and its use in verification, in Section~\ref{sec:trust}~\cite{provagent}.
In the same spirit, and because no widely used skill library yet encodes scientific procedures, we see an opportunity to package validated experimental and analysis workflows as portable, self-describing skills that move across laboratories, which would give capability discovery concrete content to advertise and reuse~\cite{agentskills}.
The original milestone M10 named specific transport protocols, and we reframe it around the MCP and A2A consolidation.
The status of M10 through M12 is summarized in Table~\ref{tab:scorecard}.

\subsection{Education and Workforce Development}
\label{sec:education}

The fifth dimension concerns preparing researchers for environments increasingly shaped by autonomous systems, where competencies span AI/ML methods, workflow thinking, human-machine collaboration, and ethical reasoning~\cite{silva2025aisle}.
The past year added empirical urgency, as a large-scale analysis of AI-engaged researchers found that they publish more and are cited more, even as the range of topics the community collectively studies contracts~\cite{aisciencefocus}.
Workforce development must therefore cultivate not only fluency with autonomous tools but also the judgment to resist their homogenizing pull.

\pp{Brief state of the art}
Scientific education still largely treats AI/ML as supplementary rather than integral, which leaves competency gaps for autonomous-laboratory settings.
National and international programs have begun to fund workforce development as an explicit component of their AI-for-science strategies~\cite{genesis,euraise}, but systematic curriculum redesign remains limited across institutions.

\pp{Challenges}
The central tension is balancing automation with foundational understanding since over-reliance risks producing scientists who cannot critically evaluate automated results.
Compounding this, current assessment methods do not measure the ability to collaborate with AI, many educators lack the relevant expertise, and access to autonomous-laboratory infrastructure for hands-on training is uneven, which raises equity concerns.

\pp{Research priorities}
We prioritize modular curricula that integrate AI/ML competencies with scientific reasoning across disciplines, immersive virtual environments that simulate autonomous laboratories where physical access is limited, and assessment methodologies that evaluate trust calibration and the interpretation of AI decisions, with ethical reasoning integrated throughout.
As agentic workflows take over routine tasks and extend what a single researcher can attempt, the design of the human-machine interface becomes a research problem in its own right.
Interfaces must let a scientist supervise, interrogate, and override an agent, calibrate trust in its output, and be augmented rather than sidelined by it.
The status of milestones M13 and M14 is summarized in Table~\ref{tab:scorecard}.

\subsection{Trust, Verification, and Reproducibility}
\label{sec:trust}

We elevate trust, verification, and reproducibility from a cross-cutting concern, as it appeared in the original roadmap, to a first-class dimension.
The motivation is the capability-reliability gap introduced in Section~\ref{sec:state}.
The ability of agents to propose hypotheses, designs, and analyses has outrun the infrastructure needed to verify them.
The past year supplied concrete evidence of the cost of this gap, from the corrected materials-discovery result we examine below~\cite{alab_correction}, to fabricated citations surfacing in papers accepted at leading venues~\cite{fabricatedcites}, to reproducibility benchmarks on which agents perform near or below chance~\cite{corebench}.
We claim that an interconnected ecosystem without commensurate verification infrastructure would amplify, not contain, these failure modes.

\pp{Brief state of the art}
Evaluation has shifted from closed-ended question answering toward reproduction and end-to-end research, with benchmarks that ask agents to reproduce published computational results~\cite{corebench} or to carry a task through the full discovery process~\cite{astabench}.
Early verification mechanisms are appearing, including uncertainty quantification, neurosymbolic constraints, and provenance that classifies each claim by its epistemic source, but they are not yet standard components of autonomous workflows.
Provenance for agentic workflows, in particular, is beginning to mature, with recent work extending the W3C PROV standard over the Model Context Protocol to capture agent prompts, responses, and decisions as near-real-time, end-to-end workflow provenance, so that an erroneous output can be traced as it propagates from one agent to the next across edge, cloud, and HPC resources~\cite{provagent}.

\pp{Challenges}
Verification in this setting is hard for four reasons.
First, the non-determinism of LLM-based orchestration is in tension with run-to-run reproducibility.
Second, fluent outputs make hallucinated results and citations harder to detect, not easier.
Third, distinguishing genuine novelty from novelty relative to a model or database requires clean train and test separation that current pipelines do not guarantee~\cite{alab_correction}.
The same difficulty recurred for a generative materials model whose experimentally highlighted compound was argued to be isostructural with a phase known for decades and already present in the model's training distribution~\cite{mattergen_critique}, which shows that the problem is not confined to a single laboratory or method.
Fourth, verifying long campaigns distributed across facilities and agents is qualitatively harder than checking a single result.

\pp{A cautionary case}
The corrected materials-discovery result illustrates the failure mode in miniature.
An autonomous laboratory reported dozens of newly synthesized compounds, and subsequent analysis showed that many were ordered versions of already-known disordered phases, that the automated structural refinement was below the quality a human expert would accept, and that one reported compound had leaked from the training data~\cite{alab_correction,alab}.
None of these errors required a flaw in the robotics.
They followed from treating an automated pipeline's output as a discovery without independent verification.
We read this not as an indictment of autonomous laboratories, but as evidence that verification must be designed in from the start, especially once results from one laboratory begin to feed the agents of another.

\pp{Research priorities}
We prioritize verification-and-validation architectures spanning the experiment lifecycle, in which physics-based and logic-based constraints act as hard boundaries rather than soft preferences, end-to-end provenance that links each claim to the data and tool executions that produced it, automated verification of citations and reported results at the point of submission~\cite{fabricatedcites}, and reproducibility metrics for autonomous workflows that make run-to-run variation measurable.
We use the two terms deliberately.
Verification asks whether a workflow was built and executed correctly, and validation asks whether its result corresponds to physical reality, the second being the harder problem in the physical sciences, where a simulation is a poor substitute for an experiment and only measurement can settle the question.
It is also why the strongest claims of autonomous discovery have so far come from mathematics and computation, where a result can be checked by a machine, rather than from physics, chemistry, or biology, where it cannot.
Because an agent can reason soundly from a mistaken picture of the world, a high-consequence action should additionally be checked against independent measurements that the agent does not control.
Two further practices sharpen this agenda.
First, the natural unit of reproducibility for an agentic workflow is the run itself, an auditable record of the plan, the tool calls, the data and model versions, the human approvals, and the evaluation outcomes that produced a result, rather than the result alone~\cite{provagent}.
That record must reach the computational execution as well, what we call \emph{AI provenance}, capturing model identities and versions, inference parameters, retrieved sources and prompt context, uncertainty estimates, and the accelerators and execution environments behind a result, so that a finding can be reproduced as models, data, and computing environments evolve.
Second, evaluation must judge the process and not only the answer, since tool choice, recovery from failure, and respect for budget and policy all bear on whether a result can be trusted~\cite{nist_airmf}.
As autonomous laboratories move toward continuous operation, efficiency itself becomes a measure of scientific productivity, so community benchmarks should report not only discovery rate and accuracy but systems-level costs, from energy per validated experiment and compute per optimization cycle to data-movement overhead, end-to-end latency, and the overhead of verification, enabling reproducible comparison across heterogeneous computing environments~\cite{greenai}.
We propose two milestones, M15 and M16, summarized in Table~\ref{tab:scorecard}.

\subsection{Safety, Security, Integrity, and Governance}
\label{sec:governance}

We add safety, security, integrity, and governance as the seventh dimension of the roadmap.
The original roadmap noted intellectual-property and liability concerns in passing, and the past year made a broader set of governance questions unavoidable.
The convergence of capable models with remotely accessible laboratories makes it easier for non-experts to run sophisticated experiments, which raises dual-use concerns that current biosecurity policy does not address~\cite{sdl_policy,dualuse_bio}.
Human interaction with self-driving labs and their LLM interfaces continues to have unexpected and unintended actions resulting in, for example, disclosure of personal information~\cite{owasp_llm02_2025}.
In parallel, the integrity of the scholarly record has come under strain, as analyses find that journal disclosure policies have done little to curb undisclosed AI-assisted writing~\cite{aipolicy}, and as community consensus holds that AI systems cannot be accountable authors.

\pp{Brief state of the art}
Technology-and-policy reviews have begun to map the governance landscape for autonomous laboratories~\cite{sdl_policy}, work on AI models and agents has started to catalog capabilities of concern~\cite{dualuse_bio,shapira2026agentschaos}, and national strategy now frames autonomous experimentation as a security as well as a scientific priority~\cite{genesis}.
Publisher norms and disclosure requirements exist, but evidence indicates that they are widely unobserved~\cite{aipolicy}, and screening at the boundary between digital design and physical execution remains largely voluntary.
Analyses of commercial cloud laboratories have made the threat concrete, observing that an experiment can be designed on a laptop in one jurisdiction and executed by robots in another, and have proposed know-your-customer screening together with a cloud-lab security consortium modeled on the existing self-regulation of commercial gene synthesis~\cite{rand_cloudlab}.

\pp{Challenges}
Dual-use risk concentrates where capable agents meet cloud and self-driving laboratories because the same automation that lowers the barrier for legitimate researchers also lowers it for misuse~\cite{sdl_policy,dualuse_bio}. 
Security, safety, and ethical concerns can arise at the human interface with agents and self-driving laboratories when there is physical proximity between humans and equipment or dangerous materials; when human-derived data is accessible to agents; or when there is the potential for misalignment of objectives and ethical norms~\cite{shapira2026agentschaos,ngong2026agentscopeevaluatingcontextualprivacy}.  
Accountability is diffuse when agents act across institutions, complicating intellectual property, liability for cross-institutional failures, and the question of who is answerable for an autonomous decision.
As campaigns increasingly span public and commercial partners, unresolved questions of data ownership and value capture become a near-term barrier rather than a distant one, and they need to be settled alongside the technical interfaces, not after them~\cite{sdl_policy}.
Throughout, governance lags capability, and a networked ecosystem widens the gap by increasing both reach and speed.
These are precisely the conditions under which a small number of failures, or a single misuse, can erode public trust in autonomous science as a whole.

\pp{Research priorities}
We prioritize human-in-the-loop guardrails with reliable override, screening, and audit at the digital-to-physical interface for networked SDLs, accountable agent identity and provenance building on Sections~\ref{sec:communication} and~\ref{sec:trust}, governance frameworks that preserve institutional autonomy while enabling collaboration, and disclosure-by-construction, in which AI-assisted contributions are recorded automatically rather than self-reported.
Trust in a distributed autonomous system also needs roots below the application layer, in trusted execution environments, cryptographically verifiable model and agent identities, signed and immutable workflow and provenance records, and hardware-rooted attestation, which together let one institution rely on another's computation without surrendering control of it.
Underlying these safeguards is a simple principle, that the autonomy granted to an agent should match the cost, risk, and reversibility of the action it would take, and should be raised only as auditable evidence of reliability accumulates rather than asserted at the outset~\cite{nist_airmf,agentsprod}.
We propose two milestones, M17 and M18, summarized in Table~\ref{tab:scorecard}.

\section{The National and Global Ecosystem}
\label{sec:ecosystem}

The original roadmap argued for a grassroots, bottom-up network on the premise that no coordinated national program existed to connect autonomous laboratories.
That premise has partly changed.
The launch of the Genesis Mission has mobilized U.S. national laboratories around an integrated platform that couples high-performance computing, scientific foundation models, datasets, and automated laboratory systems, with explicit near-term objectives for robotic laboratories and autonomous experimentation~\cite{genesis}.
In the months since launch, that mandate has acquired concrete form, as the program named twenty-six national science and technology challenges, one of which, achieving AI-driven autonomous laboratories, targets the same automated experimentation that the present roadmap addresses~\cite{genesis_challenges}.
Related efforts, such as the Trillion Parameter Consortium~\cite{tpc}, the European strategy for AI in science~\cite{euraise}, and the Acceleration Consortium~\cite{accelerationconsortium}, constitute a dense and growing institutional landscape.
That landscape now has a substantial private layer as well since the Genesis Mission enlisted two dozen industrial partners, among them chip and cloud vendors and venture-funded autonomous-discovery startups~\cite{genesis_industry}, and since those same vendors and startups are building platforms and laboratories of their own (Section~\ref{sec:state}).

Within the United States, the mobilization reaches well beyond the U.S. Department of Energy.
The National Science Foundation has established a Directorate for Technology, Innovation, and Partnerships and a milestone-funded X-Labs initiative for breakthrough science, with early topics that include scientific instrumentation~\cite{nsftip,nsfxlabs}, the Defense Advanced Research Projects Agency treating AI-driven autonomous experimentation as a national-security capability~\cite{darpabto}, and the National Institute of Standards and Technology anchoring the measurement science and data standards that autonomous laboratories require, building on the Materials Genome Initiative~\cite{nistai,mgi}.
A concrete near-term pathway is a DOE call for robotics and automation testbeds for autonomous scientific discovery, framed as reusable community infrastructure for the national laboratories and their industry partners~\cite{ascrtestbeds}.
Closest of all to this roadmap's own premise, a National Science Foundation test-bed program aims to build a network of programmable cloud laboratories, remotely accessible autonomous facilities to be linked by computational networking and shared data and AI standards~\cite{nsfpcl}.
This multi-agency picture strengthens, rather than weakens, the case for a coordinating fabric because each program risks building its own island unless interfaces and standards are held in common.

The European setting shows how much this division of labor depends on local conditions.
Public compute for large-scale AI is concentrated in a few sites, such as the EuroHPC exascale system JUPITER at J\"ulich and the national centres of the Gauss Centre for Supercomputing, while compute at the laboratories themselves is more modest, and high-speed networking between laboratories and compute sites is not yet in place~\cite{eurohpc}.
Federation across facilities, which much of this roadmap presumes, is therefore gated by infrastructure that is unevenly available.
Access to the most widely used proprietary models is likewise not guaranteed in every region, and European alternatives remain sparse, which is one more reason that open-weight models with transparent provenance (Section~\ref{sec:orchestration}) matter beyond reproducibility alone.
We include this less as a digression than as a reminder that a roadmap written largely against U.S. mobilization must be read and adapted against the compute, data, and network realities of each region that adopts it.

We argue that top-down mobilization and bottom-up coordination are complementary rather than redundant.
Large programs supply what a grassroots network cannot, including compute at scale, foundation models, a security framing, and funding at scale~\cite{genesis,tpc}.
A grassroots network, in turn, supplies what large programs and commercial vendors tend to underweight, namely vendor-neutral interfaces and cross-institutional standards that prevent each new program or proprietary platform from becoming another silo, together with the deliberate inclusion of resource-constrained institutions through portable, low-footprint laboratory modules.

In this division of labor, an interconnected network such as AISLE is best understood not as an alternative to federal mobilization but as the interoperability fabric and community-standards layer that allows these initiatives to connect to one another and to the broader research community, rather than to interconnect only internally.
We therefore see the roadmap of this paper as a contribution that is independent of but synergistic with the national and global programs now taking shape.
A practical consequence is that the milestones we revise below should be read as community commitments that any of these programs can adopt, instrument, and report against, rather than as the agenda of a single institution.
As concrete evidence that such adoption is feasible, the Genesis Mission consortium has organized its own work into groups for robotics and automation, data integration and standards, model development and validation, and computing infrastructure, which align respectively with the instrument, data, trust, and orchestration dimensions of Section~\ref{sec:dims}~\cite{genesis_consortium}.

\section{Conclusion and Revised Roadmap}
\label{sec:conc}

In this paper, we presented an updated community roadmap for interconnected autonomous science, one year after the original AISLE roadmap~\cite{silva2025aisle}.
We characterized seven shifts in the landscape, namely validated discoveries, a second generation of self-driving laboratories, a consolidating interface layer, reasoning and foundation models as a new substrate, benchmarks that expose a capability-reliability gap, the move of trust and governance to the foreground, and the entry of industry as a primary actor. We used two lenses, an autonomy ladder and that same gap, to interpret them.
We refined the five original dimensions in light of this evidence, and we elevated two concerns, trust and verification (Section~\ref{sec:trust}) and safety, security, and governance (Section~\ref{sec:governance}), to first-class dimensions of the roadmap.
Our milestone-by-milestone assessment is collected in the scorecard of Table~\ref{tab:scorecard}, presented with the dimensions in Section~\ref{sec:dims}.
The pattern is consistent in that the dimensions closest to raw model capability have advanced the fastest, while those that require cross-institutional infrastructure, verification, and governance remain largely open.
Orchestration (Section~\ref{sec:orchestration}) and agent interfaces (Section~\ref{sec:communication}) moved the furthest; the former carried by reasoning and foundation models and the latter by the consolidation of agent protocols.
Data management, education, and the federated aspects of instrument integration moved the least because they depend on coordination that no single model improvement can supply.
The two dimensions we add, trust and governance, do not appear on the original scorecard precisely because the past year revealed them to be prerequisites rather than refinements.
Read as a two-year plan, the first year concentrates on interfaces, protocol adoption, AI-driven metadata, and the scaffolding of verification, while the second year targets federation, zero-trust coordination, cross-facility knowledge integration, and governance.
We deliberately keep the horizon short because, at the current pace of change, a longer-range plan would be obsolete before it could be acted upon.

In future work, we plan to convene the community around this two-year roadmap to develop reference implementations that exercise the consolidated interface layer (Section~\ref{sec:communication}) across multiple facilities and to prototype the verification, validation, and governance mechanisms that the past year has shown to be prerequisites, rather than refinements, for trustworthy autonomous discovery.

% References
\newpage
\cleardoublepage\phantomsection\addcontentsline{toc}{section}{References}
\bibliographystyle{IEEEtranDOI}
{\small \bibliography{refs}}

\end{document}